\documentstyle[psfig,iopconf1,12pt]{article}
\def\zp{\mbox{$Z^\prime$ }}
\def\nim#1#2#3{           { Nucl. Inst. \& Meth. }{\bf #1}, #3 (19#2)}
\begin{document}

\title{ Neutrino electron scattering and electroweak gauge structure:
Probing the masses of a new Z boson}
\author{O. G. Miranda\dag
\footnote{On leave from Departamento de F\'{\i}sica 
CINVESTAV-IPN, A. P. 14-740, M\'exico 07000, D. F., M\'exico}, 
V. B. Semikoz\ddag
Jos\'e W. F. Valle\dag \footnote{E-mail: valle@flamenco.ific.uv.es}}

\affil{\dag Instituto de F\'{\i}sica Corpuscular 
- C.S.I.C.\\Departament de F\'{\i}sica Te\`orica, Universitat de
Val\`encia\\46100 Burjassot, Val\`encia, Spain\\
http://neutrinos.uv.es}
\affil{\ddag Institute of the 
Terrestrial Magnetism, the Ionosphere and Radio Wave Propagation of the 
Russian Academy of Science, IZMIRAN, Troitsk, Moscow region, 142092, Russia.
}
\beginabstract 

Low-energy high-resolution neutrino-electron scattering experiments
may play an important role in testing the gauge structure of the
electroweak interaction. We discuss the use of strong radioactive
neutrino sources (e.g. $^{51}$Cr) in underground experiments such as
BOREXINO, HELLAZ and LAMA. We display the sensitivity of these
detectors in testing the possible existence of extra neutral $E_6$
gauge bosons.

\endabstract

\section{Introduction}
\vskip .1cm 

Despite the success of the Standard Model (SM) in describing the
electroweak interaction, there have been considerable interest in
extensions of the gauge structure of the theory~\cite{fae}.  In this
talk we discuss the proposal~\cite{msv} of using $\nu_{e} e $ and
$\overline{\nu_{e}} e$ scattering from terrestrial neutrino sources
with improved statistics as a test of the electroweak gauge structure.
The coupling constants governing $\nu_e e\to \nu_e e$ scattering in
the SM have been well measured from $e^{+}e^{-} \to l^{+}l^{-}$ at
high energies by the LEP Collaborations and have given strong
constraints on additional neutral currents, specially on the mixing of
the standard Z boson with other hypothetical neutral gauge bosons.
This carries an important weight in global fits of the electroweak
data~\cite{lang}. However, we argue that low-energy experiments can
give complementary information, namely, they allow a better
sensitivity to the mass of the new gauge boson than available from
high energies, e.g. from LEP physics. On the other hand, although the
Tevatron does give relatively good limits on \zp masses, one may argue
that a neutrino-electron experiment is a cleaner environment that will
provide useful complementary information on the gauge structure of the
electroweak interaction.

Using $\nu_{e} e $ and $\overline{\nu_{e}} e$ scattering from
terrestrial neutrino sources has also been suggested as a test of
non-standard neutrino electromagnetic properties, such as magnetic
moments~\cite{Vogel}.  In contrast to reactor experiments such MUNU
\cite{MUNU}, a small (but intense) radioactive isotope source can be
surrounded by detectors with full geometrical coverage.  Here we
demonstrate that a low-energy high-resolution experiment can play an
important role in testing the structure of the neutral current weak
interaction.

We explicitly determine the sensitivity of these radioactive neutrino
source experiments as precision probes of the gauge structure of the
electroweak interaction and illustrate how it works in a class of
$E_6$-type models as well as models with left-right symmetry.

\section{The $\nu e$ Cross Section}

In a generic electroweak gauge model the differential cross section
for the process $\nu_e e\to \nu_e e$ is given by

\begin{eqnarray}
\frac{d\sigma}{dT}  = 
 \frac{2m_{e}G^2_{F}}{\pi} \big\{ 
    (g_{L}+1)^2+ g_{R}^2 (1-\frac{T}{E_{\nu}})^2  - 
   \frac{m_e}{E_{\nu}} (g_{L}+1)
    g_{R}\frac{T}{E_{\nu}} \big\}  \label{DCS}
\end{eqnarray}
where $T$ is the recoil electron energy, and $E_{\nu}$ is the neutrino
energy.  For the SM case we have $g_{L,R}=\frac12 (g_{V}\pm g_{A})$,
with $g_{V}=\rho_{\nu e}(-1/2+2\kappa \mbox{sin}^2\theta_{W})$ and
$g_{A}=-\rho_{\nu e}/2$ where $\rho_{\nu e}$ and $\kappa$ describe the
radiative corrections for low-energy $\nu_{e} e\to\nu_{e} e $
scattering~\cite{Sirlin}. For the case of $\overline{\nu_{e}} e \to
\overline{\nu_{e}} e$ scattering we just need to exchange $g_{L}+1$
with $g_{R}$ and vice versa.

As already mentioned, the values of the coupling constants governing
$\nu_e e\to \nu_e e$ scattering in the SM have been well measured
through the $e^{+}e^{-} \to l^{+}l^{-}$ process at LEP.  A combined
LEP fit at the Z peak gives~\cite{Stickland} $g_{V}=-0.03805\pm
0.00059$ and $g_{A}=-0.50098 \pm 0.00033$.  These results have given
strong constraints on the mixing of the standard Z boson with an
additional \zp, in the framework of global fits of the electroweak
data~\cite{lang}. As a result we will, in what follows, focus mainly
on the possibility of probing the \zp mass at low-energy $\nu_e e\to
\nu_e e$ scattering experiments. For convenience we define the
parameter
\begin{equation}
\gamma=\frac{M^2_{Z}}{M^2_{Z^\prime}}
\end{equation}
and we will neglect the mixing angle $\theta^\prime$ between the SM
boson and the extra neutral gauge boson.

For extended models, the neutral contribution to the differential 
cross section will be, for $\theta^\prime=0$, 
\begin{equation}
\delta\frac{d\sigma}{dT} =\gamma\Delta = \gamma\frac{2m_{e}G^2_{F}}{\pi}\big\{ 
    D  + E \frac{T}{E_{\nu}}(\frac{T}{E_{\nu}}-2) - 
    F \frac{m_e}{E_{\nu}} \frac{T}{E_{\nu}}\big\} \label{dc}
\end{equation}
with $\Delta$ in obvious notation and 
\begin{eqnarray}
D&=&2(g_{L}+1)\delta g_{L}+ 2g_{R} \delta g_{R} \\
E&=&2g_{R}\delta g_{R} \\
F&=& (g_{L}+1)\delta g_{R}+
    g_{R}\delta g_{L}  \label{coe}
\end{eqnarray}
where $g_{L}$ and $g_{R}$ are the SM model expressions and $\delta
g_{L,R}$ give the corrections due to new physics.  In the particular
case of the LRSM these corrections are given by \cite{LR1,LR2}
\begin{eqnarray}
\delta g_{L} &=& \frac{s^4_{W}}{r^2_{W}}g_{L} 
                 +\frac{s^2_{W}c^2_{W}}{r^2_{W}}g_{R}  \\
\delta g_{R} &=&\frac{s^4_{W}}{r^2_{W}}g_{R} 
                 +\frac{s^2_{W}c^2_{W}}{r^2_{W}}g_{L} 
\end{eqnarray}
while for the $E_6$ models we have \cite{npb345,Concha}, 
\begin{eqnarray}
\delta g_{L} &=& 4\rho s^2_{W} (\frac{3\mbox{cos}\beta}{2\sqrt{24}}
        +\frac{\sqrt{5}}{\sqrt{8}}\frac{\mbox{sin}\beta}{6})
        (\frac{3\mbox{cos}\beta}{\sqrt{24}}
         +\frac13 \frac{\sqrt{5}}{\sqrt{8}}\mbox{sin} \beta)   \\
\delta g_{R} &=& 4\rho s^2_{W} (\frac{3\mbox{cos}\beta}{2\sqrt{24}}
        +\frac{\sqrt{5}}{\sqrt{8}}\frac{\mbox{sin}\beta}{6})
        (\frac{\mbox{cos}\beta}{\sqrt{24}}
         -\frac13 \frac{\sqrt{5}}{\sqrt{8}}\mbox{sin}\beta) 
\end{eqnarray}
where, $\rho$ includes the radiative corrections to the ratio 
$M^2_{W}/M^2_{Z}cos\theta_{W}$ and $\beta$ defines the $E_6$ model in which we 
are interested in. 

The correction to the $\nu_{e} e$ scattering depends on the model as
well as on the energy region. In order to illustrate how this
corrections affect the SM prediction we define the expression

\begin{equation}
R=\frac{\Delta}{(\frac{d\sigma}{dT})^{SM}}.\label{rat}
\end{equation}
This ratio depends on the specific model through the angle $\beta$ and
depends also on the electron recoil energy, as well as on the neutrino
energy. As we are interested in artificial neutrino sources we can
study what would be the value of $R$ in Eq. (\ref{rat}) for the case
of a neutrino coming from a $^{51}$~Cr source, which corresponds to a
neutrino energy $E_{\nu}=746$ KeV.
We show this ratio in Fig. 1 as a function of $\beta$ for different
values of $T$.  We can see from the plot that the sensitivity is
bigger at $\mbox{cos}\beta \simeq 0.8$ and it is almost zero for
$\mbox{cos}\beta\simeq -0.4$. Of the most popular models ($\chi$,
$\eta$ and $\psi$ models) we can say that the $\chi$ model is the most
sensitive to this scattering.  A similar result can be obtained for
the case of anti-neutrino sources, such as $^{147}$~Pm, now proposed
for the LAMA experiment \cite{Barabanov,lama}.  We can also see from
the figure that, in order to reach a constraint on $\gamma \simeq.1$
in the $\chi$ model we need a resolution of the order of 5~\%.

\begin{figure}
\centerline{\protect\hbox{\psfig{file=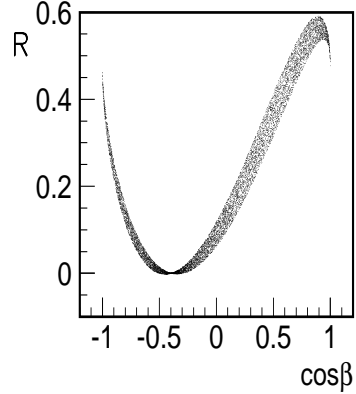,height=6.cm,width=5cm}}}
\caption{Plot of the ratio given in eq. \protect{\ref{rat}} as a
function of the model for different values of $T$ and for $E_{\nu}=746$ 
KeV.}
\end{figure}

\section{Experimental prospects}

The first high-activity artificial neutrino sources have been recently
developed in order to calibrate both GALLEX and SAGE solar neutrino
experiments \cite{GALLEX}. These are $^{51}Cr$ sources producing
neutrinos by electron capture through the reaction $^{51}~Cr+e \to
^{51}~V+\nu_{e}$. The main line is at 746 KeV and represents 90 \% of
the neutrino flux. Besides the neutrino flux, there is also $\gamma$
emission which is stopped by a tungsten shielding. The activity of the
GALLEX source was 1.67 $\pm 0.03$ MCi.

An anti-neutrino source has recently been considered by the LAMA
proposal in order to probe for the neutrino magnetic moment
\cite{Barabanov}.  This is a $^{147}Pm$ source that produces
anti-neutrinos through the $^{147}Pm \to ^{147}Sm+e+
\overline{\nu_{e}}$ beta decay. In this case we have an anti-neutrino
spectrum with energies up to 235 KeV. The spectrum shape is well known
and the activity of the source can be measured with an accuracy better
than 1 \% \cite{Kornoukhov}. A tungsten shielding of 20 cm radius plus
a Cu shielding of 5 cm is considered in order to stop the $\gamma$
emission. In this case we can use as a good approximation for the
anti-neutrino spectrum the expression

\begin{figure}
\centerline{\protect\hbox{\psfig{file=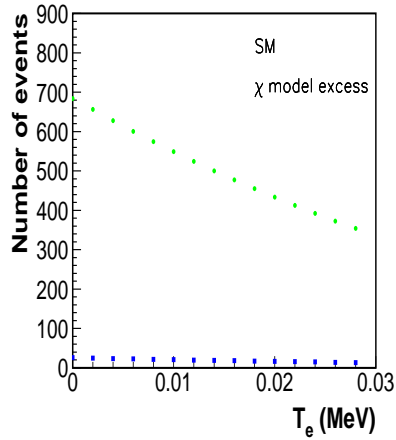,height=6.cm,width=5cm}}}
\caption{Expected number of events per bin for the LAMA proposal. The 
additional contribution of an extra neutral gauge boson with 
$M_{Z^\prime}=330$~GeV in the $\chi$ model is also shown. }
\end{figure}

\begin{equation}
f(E_{\nu} )  = \frac{1}{N} F(Z,p) E_{\nu}^2 (W-E_{\nu})
 \sqrt{(W-E_\nu)^2-m_e^2} 
\end{equation}
where  $F(Z,p)$ stands for the Coulomb correction to the spectrum,  
$N$ is a normalisation factor and $W=m_e+235$~KeV. 

The possibility of surrounding the source with a NaI(Tl) detector is
now considered by the LAMA team. As a first step they plan to use a
400 tones detector (approximately $2\times 10^{29}$ electrons) that
will measure the electron recoil energy from 2 - 30 KeV; the source
activity will be 5 MCi. A second stage with a one tone detector and 15
MCi of $^{147}Pm$ is under study.

We can now estimate the event rates expected both in the Standard
Model as well as in extended models for the configuration discussed
above. In order to do this we need to integrate over the neutrino
energy spectrum and to take the average over the electron recoil
energy resolution. The expected number of events per bin in the
Standard Model is shown in Fig. 2. For definiteness we have considered
2 KeV width bins. In Fig. 2 we also show the excess in the number of
events for the case of an extra neutral gauge boson in the $\chi$
model for a $Z^\prime$ mass of 330 GeV.  The prospects of the
experiment to be sensitive to such an excess in the shape of the
electron energy distribution will depend on the error achieved in the
event numbers per bin. At the moment we can only estimate the
statistical error, but not for the systematic.

In order to estimate the LAMA sensitivity to the mass of a $Z^\prime$
in the $\chi$ model we have considered an experimental set up with 5
MCi source and a one tone detector. Assuming that the detector will
measure exactly the SM prediction and taking into account only the
statistical error, we obtain a sensitivity of the order of 600 GeV at
95 \% C. L., comparable to the present Tevatron result. A more
detailed analysis can be found in ref.~\cite{lama}.

In the case of the BOREXINO proposal \cite{BOREXINO}, they have
considered the use of a $^{51}Cr$ source that will be located at 10 m
from the detector \cite{Fiorentini}; unfortunately the experimental
set up does not allow one to surround the source and, therefore the
statistics is not high enough to provide a strong sensitivity to the
\zp mass. In this case, if we consider again that the experiment will
measure the SM prediction, the sensitivity to the $Z^\prime$ mass in
the $\chi$ model will be 275 GeV, if only the statistical error is
considered. If we take into account the background \cite{Giammarchi}
the sensitivity will decrease to 215 GeV.

\begin{figure}
\centerline{\protect\hbox{\psfig{file=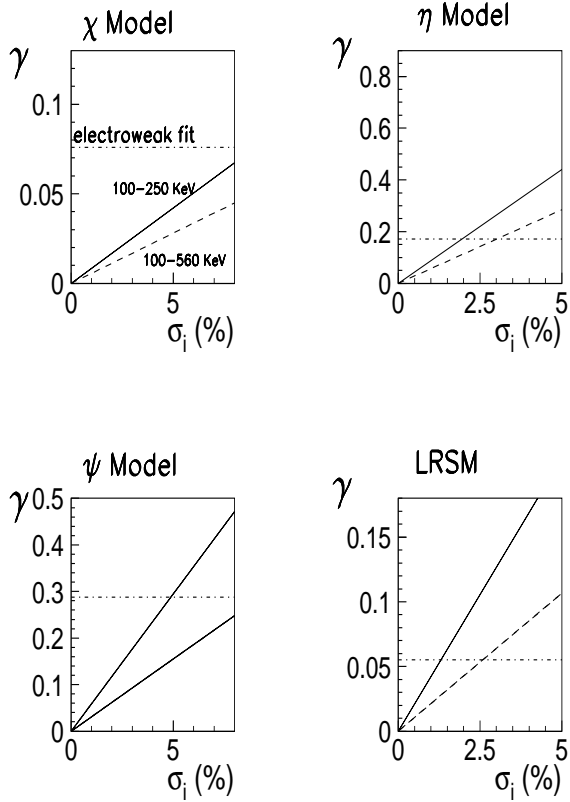,height=12.cm,width=8cm}}}
\caption{ Sensitivity to the $\gamma$ parameter for different models
for the case of the HELLAZ proposal. The solid line correspond to the
case of an energy region from 100 KeV - 250 KeV while the dotted line
is for the region from 100 KeV - 550 KeV. We also show in the plot the 
present constraint from indirect searches \cite{lang}}
\end{figure}

Finally we now move to the HELLAZ proposal. Although this
collaboration has not considered the use of an artificial source,
there is room to speculate about the experimental set up and expected
event rates. For definiteness we assume in our analysis a $^{51}Cr$
source and the originally designed HELLAZ detector \cite{HELLAZ}.
Since the error will depend on the specific topology of the
experiment, we have computed the sensitivity at 95 \% C. L. for
different values of the total error in the number of events per bin (a
detailed explanation of this analysis can be found in \cite{msv}). The
results, for four different models are shown in Fig. 3 where we have
considered two possible energy regions for the detector. First we
consider the case of an energy window from 100 KeV-250 KeV, that is
the energy region that HELLAZ is considering for the study of solar
neutrinos. We can see that the prospects for getting a better
sensitivity than other indirect searches seems to be very realistic.
On the other hand the chances of improving the Tevatron constraint
seems feasible only if a high-statistics experiment is done. This can
be achieved either by constructing a more intense source, by
increasing the mass of the detector, by enriching the source several
times (in order to increase the exposure time), or a combination of
the above. Extending the energy window to 100 KeV-560 KeV is also
helpful in getting a better sensitivity, as can be seen from the same
Fig 3.

\noindent {\bf Acknowledgements} 

We would like to thank useful discussions with Igor Barabanov, Marco
Giammarchi, Concha Gonz\'alez-Garc\'{\i}a, and Tom Ypsilantis.  This
work was supported by DGICYT under grant number PB95-1077, by the TMR
network grant ERBFMRXCT960090 and by INTAS grant 96-0659 of the
European Union. O. G. M. was supported by a CONACYT and SNI-M\'exico,
and V. S. by the sabbatical grant SAB95-506 and RFFR grants
97-02-16501, 95-02-03724.

\end{document}